\documentclass{article}

\usepackage{amsmath}
  \usepackage{amssymb}
  \usepackage{mathrsfs}
\usepackage{epsf}
\usepackage{epsfig}

\newcommand{\be}{\begin{equation}}
\newcommand{\ee}{\end{equation}}
\newcommand{\bea}{\begin{eqnarray}}
\newcommand{\eea}{\end{eqnarray}}
\newcommand{\bg}{\begin{gather}}
\newcommand{\eg}{\end{gather}}

\newcommand{\bb}{\bibitem}
 
\newcommand{\eqn}{\begin{eqnarray}}
\newcommand{\eqnx}{\end{eqnarray}}

\begin{document}

\date{} 

\title{On Bogomolny equations in the Skyrme model}

\author{{\L}. T. St\c{e}pie\'{n}\footnote{The Pedagogical University of Cracow, ul. Podchora\c{}\.{z}ych 2, 30-084 Krak\'{o}w, Poland,
e-mail address: sfstepie@cyf-kr.edu.pl, lukasz.stepien@up.krakow.pl}}

\maketitle

\begin{abstract}
Using the  concept of strong necessary conditions (CSNC), we derive a complete decomposition of the minimal Skyrme model into a sum of three coupled BPS submodels with the same topological bound. The bounds are saturated if corresponding Bogomolny equations, different for each submodel, are obeyed. 
\end{abstract}

\section{Introduction}

BPS models are classical field theories (1) admitting reduction of the full second order static equations of motion to a set of first order equations (so-called Bogomolny or BPS equations 
\cite{BPS}-\cite{PraSom}) which solutions (2) saturate a pertinent topological lower bound on the static energy. 

BPS  models play an important role in physics. Owing to Bogomolny equations, obtaining of exact solutions is possible. Such solutions significantly enlarge our understanding of considered non-linear models. In fact, one may treat BPS models as "harmonic oscillators" of nonlinear classical field theories (with a topological charge) where many questions can find analytical and exact answers. Moreover, due to the saturation of a pertinent energy bound, solutions Bogomolny equations are necessary the lowest energy states in each topological sector which guarantees the topological stability of solitons carrying a non-trivial value of the corresponding topological charge. 

Therefore, models with the BPS property are wanted. There exist several  methods of derivation of BPS equations: first of them is based on the original Bogomolny trick i.e., completing to a square \cite{BPS}-\cite{Boggen}. Some other approaches are: the first order formalism \cite{bazeia} and on-shell method \cite{onshell}). However, a completely general method, which allows for a systematic derivation (if possible) BPS equations, is called as  {\it the concept of strong necessary conditions} (CSNC). It was originally introduced and analyzed in \cite{sokalski1979}-\cite{stepien2012_2018}, and it has been very recently further developed by Adam and Santamaria \cite{adam}, who proposed so called {\it first order Euler-Lagrange} (FOEL) formalism.  

In this paper, we apply the CSNC method to derive a {\it complete} BPS structure of the generalized Skyrme model \cite{skyrme}, which is widely considered as a candidate for a low energy limit of QCD which has an ability to describe all baryonic (colorless) states in the nature - from single baryons and atomic nuclei to nuclear matter and neutron stars. In other words, we accomplish the program started recently in \cite{submodels}, \cite{submodels2}. 

\section{The Skyrme model}
The generalized $SU(2)$ Skyrme model is defined by the following Lagrange density 
 \begin{equation}
 \mathcal{L} = {\mathcal{L}}_{0} + {\mathcal{L}}_{2} + {\mathcal{L}}_{4} + {\mathcal{L}}_{6},
 \end{equation}
where we have a two derivative term (kinetic or Dirichlet term)
	\begin{equation} 
  \mathcal{L}_{2} =- \frac{1}{2} Tr(L_{\mu} L^{\mu}),
	\end{equation} 
	a four derivative term (Skyrme term)
	\begin{equation} 
 \mathcal{L}_{4} = \frac{1}{16} Tr([L_{\mu}, L_{\nu}]^{2}),
	\end{equation} 
	a six derivative term (sometimes referred as the BPS term)
	\begin{equation} 
  \mathcal{L}_{6} = \lambda^{2} \pi^{2} \mathcal{B}_{\mu} \mathcal{B}^{\mu},
	\end{equation} 
where 
$$\mathcal{B}^{\mu} = \frac{1}{24 \pi^{2}} \varepsilon^{\mu \nu \rho \sigma}Tr(L_{\nu} L_{\rho} L_{\sigma})$$
is the topological (baryonic) current. 
Here $L_{\mu} = U^{\dagger} \partial_{\mu} U$ is the left invariant current and $U\in SU(2)$ is the Skyrme matrix field. Finally we have a non-derivative term, that is a potential $\mathcal{L}_{0} = -m^{2} \mathcal{V}(Tr(U))$, where $m$ can be related to a mass of small perturbations (pions). Note, that for the first terms in the Lagrangian we omit the usual coupling constants $f_\pi$ and $e$ which can be re-introduced by a suitable change of length and energy units. 
	
	At the beginning, let us remind that the generalised Skyrme model is not an example of a BPS model. It is a consequence of the fact that even the minimal Skyrme model i.e., $\mathcal{L}_{24}=\mathcal{L}_2+\mathcal{L}_4$ does not possesses nontrivial solutions saturating a corresponding topological bound, so-called Faddeev bound \cite{F}, \cite{MS}
\be
E \geq 12\pi^2 |B|
\ee	
However, there is a rather reach BPS structure hidden in the full model. First of all, there is BPS submodel, referred as the BPS Skyrme model, $\mathcal{L}_{BPS}= \mathcal{L}_6+\mathcal{L}_0$ which {\it is} a genuine BPS theory with a  topological bound saturated by infinitely many solitons (BPS Skyrmions) in an arbitrary topological sector \cite{BPS skyrme}. An importance of this finding is related to a problem of too higher binding energies in the original $\mathcal{L}_{24}$ model. The BPS theory has necessary zero classical binding energies while small contributions can show up due to semiclassical quantisation and inclusion of the Coulomb interactions \cite{nuclei}. 
As the BPS Skyrme model is a point in the parameter space of the full model (i.e., a limit where coefficients multiplying $\mathcal{L}_2$ and $\mathcal{L}_4$ vanish) one can use it as a starting point for a whole family of near-BPS Skyrme type models with physically small classical binding energies. (For other Skyrme type model saturating a BPS bound see \cite{other}.) 

Secondly, even the minimal part $\mathcal{L}_{24}$ enjoys an interesting BPS structure  \cite{submodels},  \cite{submodels2}.  To see this we need to introduce explicit coordinates on $SU(2) \cong \mathbb{S}^3$. Specifically we use the standard parametrization of the $SU(2)$ field $U$ by one real scalar $\xi$ and one three component isovector $\vec{n}$ of unit length
	\begin{equation}
	  U = \exp{(i \xi \vec{\tau} \cdot \vec{n})},
	\end{equation}
where $\vec{\tau}$ are the Pauli matrices. Furthermore, $\vec{n}$ can be expressed by a complex scalar $\omega$ by the stereographic projection $$\vec{n} = \left[\frac{\omega + \omega^{\ast}}{1 + \omega \omega^{\ast}}, \frac{-i(\omega + \omega^{\ast})}{1 + \omega \omega^{\ast}}, \frac{1 - \omega \omega^{\ast}}{1 +  
  	\omega \omega^{\ast}} \right]. $$
Following \cite{submodels} and \cite{submodels2} we write the two parts of the $\mathcal{L}_{24}$ Skyrme model as 
			\begin{eqnarray}
			\mathcal{L}_{2} = \mathcal{L}^{(1)}_{2} + \mathcal{L}^{(2)}_{2}, \  \mathcal{L}^{(1)}_{2} = 4 \frac{\sin^{2}{(\xi)}}{(1 + \omega \omega^{\ast})^{2}} \omega_{\mu} \omega^{\ast \mu}, \ \mathcal{L}^{(2)}_{2} = \xi_{\mu} \xi^{\mu}, \\
			\mathcal{L}_{4} = \mathcal{L}^{(1)}_{4} + \mathcal{L}^{(2)}_{4}, \ \mathcal{L}^{(1)}_{4} = 4 \sin^{2}{(\xi)} \bigg( \xi_{\mu} \xi^{\mu} \frac{\omega_{\mu} \omega^{\ast \mu}}{(1 + \omega \omega^{\ast})^{2}} - 
			\frac{\xi_{\mu} \omega^{\ast \mu} \xi_{\nu} \omega^{\nu}}{(1 + \omega \omega^{\ast})^{2}} \bigg), \\
			\mathcal{L}^{(2)}_{4} = 4 \sin^{4}{(\xi)} \frac{(\omega_{\mu} \omega^{\ast\mu})^{2} - \omega^{2}_{\mu} \omega^{\ast 2}_{\nu}}{(1 + \omega \omega^{\ast})^{2}}. 
		  \end{eqnarray} 
Therefore, 
\be
\mathcal{L}_{24}=\left(\mathcal{L}^{(1)}_2 +  \mathcal{L}_{4}^{(1)} \right)+\left(\mathcal{L}^{(2)}_2 +  \mathcal{L}_{4}^{(2)} \right) \equiv \mathcal{L}^{(1)} +\mathcal{L}^{(2)}
\ee
where each of the constituent submodels $\mathcal{L}^{(1)}, \mathcal{L}^{(2)}$, if taken separately, is a proper BPS model. Indeed, the static energy of the first BPS submodel can be written as
\be
E^{(1)}= \int d^3 x \frac{4\sin^2 \xi}{(1 + \omega \omega^{\ast})^2} \left[ \omega_i \omega_i^{\ast}+ \xi_j^2(\omega_i \omega_i^{\ast}) - (\xi_i \omega_i ) (\xi_j \omega_j^{\ast} ) \right] \geq 8\pi^{2} |B|,
\ee
where
\begin{equation}
\begin{gathered}
B= \int B_{0} d^{3} x =  \frac{1}{\pi^2} \int d^3 x \frac{i \sin^{2}{\xi}}{(1 + f^2)^{2}} \varepsilon_{ijk} \xi_{i} \omega_{j} \omega^{\ast}_{k} \end{gathered} 
\end{equation}
The bound is saturated for solution of the following Bogomolny equation 
\be
 \omega_{i} \pm i \varepsilon_{ijk} \xi_{j} \omega_{k} = 0 \label{bog 1}
\ee
and its complex conjugation. 

Analogously, the static energy for the second BPS submodel is
\be
E^{(2)}= \int d^3 x \left( \xi_i^2 + 4\sin^4 \xi \frac{1}{(1+|u|^2)^4} (i\epsilon_{ijk} u_j \bar{u}_k)^2 \right)\geq  4\pi^2 |B| 
\ee
where the corresponding Bogomolny equation is 
\be
\xi_i \mp  \frac{2i\sin^2 \xi}{(1 + \omega \omega^{\ast})^2} \epsilon_{ijk} \omega_j \ \omega^{\ast}_k =0 \label{bog 2}
\ee
Together, both bounds provide the Faddeev bound. Moreover, there is no common solutions of these Bogomolny equations in $\mathbb{R}^3$ base space \cite{manton}. 

Note, that the number of the independent equations resulting from the Bogomolny equations for the first BPS submodel (\ref{bog 1}) is twice as  for the second BPS submodel (\ref{bog 2}). As a result, the Bogomolny equations of the first submodel can be expressed as 
\be
\lambda_2 = \pm \lambda_1 \lambda_3 \;\;\; \mbox{and} \;\;\; \lambda_3 = \pm \lambda_1 \lambda_2
\ee
while for the second submodel they are equivalent to 
\be
\lambda_1 = \pm \lambda_2 \lambda_3
\ee
where $\lambda_i^2$ are the eigenvalues of the strain tensor $D_{ij}=-\frac{1}{2} Tr (L_iL_j)$ \cite{MS}, \cite{manton}. Such a nonequivalence in number of independent equations for the Bogomolny equations (\ref{bog 1}) and  (\ref{bog 2}) leads to a question if it is possible to further decompose the $\mathcal{L}_{24}$ Skyrme to a collection of {\it three} BPS submodels such that each of them corresponds to one real scalar equation (related to one constrain on the eigenvalues $\lambda_i$). Then, each submodel would contribute to the total Faddeev bound in the same way.

Obviously, to find such a complete BPS structure of the $\mathcal{L}_{24}$ Skyrme model it is enough to consider the first BPS submodel. Then we further decompose the complex field by two real degrees of freedom $\omega = f e^{ig}$. Hence, 
\bea E^{(1)} &=& 4 \int \frac{\sin^{2}{\xi}}{(1 + f^2)^{2}} [ f^{2}_{i} + f^{2} g^{2}_{i} + f^{2} (\varepsilon_{ijk} \xi_{j} g_{k})^{2} +
(\varepsilon_{ijk} \xi_{j} f_{k})^{2} ] d^{3} x\\ &= & E^{(1)}_{1} + E^{(1)}_{2}, 
\eea
where
 \begin{equation}
\begin{gathered}
E^{(1)}_{1} = 4 \int \frac{\sin^{2}{\xi}}{(1 + f^2)^{2}} [ f^{2}_{i} + f^{2} (\varepsilon_{ijk} \xi_{j} g_{k})^{2} ] d^{3} x
  \label{first_subsub}
\end{gathered} 
\end{equation}
and
\begin{equation}
\begin{gathered}
E^{(1)}_{2} = 4 \int \frac{\sin^{2}{\xi}}{(1 + f^2)^{2}} [ f^{2} g^{2}_{i} + (\varepsilon_{ijk} \xi_{j} f_{k})^{2} ] d^{3} x
 \label{second_subsub}  
\end{gathered} 
\end{equation}

In the subsequent analysis we will investigate the existence and properties of Bogomolny equations for these new subsectors of the $\mathcal{L}^{(1)}$ theory (i.e., the first BPS Skyrme submodel). To accomplish this aim we will apply the CSNC method. So, we begin with a short summary on this approach. 

\section{The concept of strong necessary conditions}

The main idea of the concept of strong necessary conditions (shortly: CSNC), is such that instead of studying the Euler-Lagrange equations:

 \begin{equation}
 \frac{d}{d x^{\mu}} \mathcal{L}_{,\Phi^{\nu}_{,x^{\mu}}} 
- \mathcal{L}_{,\Phi^{\nu}} = 0, \label{r_EL}
\end{equation}
 we study the differential equations, generated by the strong necessary conditions, \cite{sokalski1979} - \cite{SWL3}.
\begin{gather}
\mathcal{L}_{,\Phi^{\nu}_{,x^{\mu}}} = 0, \label{rd1} \\ 
\mathcal{L}_{,\Phi^{\nu}} = 0. \label{rd2}
\end{gather} 
Obviously, the set of the solutions of these equations, is a subset of the set of the solutions of the equations 
(\ref{r_EL}). However, very often one can obtain rather trivial solutions of the equations (\ref{rd1}) - (\ref{rd2}). On the other hand,  
we can prevent to it, by doing the following gauge transformation, \cite{sokalski1979} - \cite{SWL3}
\begin{equation}
\mathcal{L} \Longrightarrow \mathcal{L} + I, \label{gaugedlagr}
\end{equation}
where $I$ is such functional that $\delta I \equiv 0$.

After applying the strong necessary conditions (\ref{rd1}) - (\ref{rd2}), for the gauged Lagrangian (\ref{gaugedlagr}), we get 
dual equations \cite{sokalski1979} - \cite{SWL3}. As one can see, the Euler-Lagrange equations (\ref{r_EL}) are invariant with respect to the gauge transformation (\ref{gaugedlagr}), but the strong necessary conditions 
(\ref{rd1}) - (\ref{rd2}) are not invariant with respect to this transformation. Hence, we can extend the set of the solutions of the strong necessary conditions.  
Next, in order to obtain Bogomolny decomposition (Bogomolny equations, Bogomol'nyi equations), we need to make the dual equations self-consistent, \cite{sokalski2002} - \cite{sokalski2009}. This will be shown in the next sections, how one can do it, in the cases investigated in this paper.

\section{DECOMPOSITION OF THE FIRST BPS SUBMODEL}

\subsection{First subsubmodel}

We derive strong necessary conditions for (\ref{first_subsub}) with a generalization:
\begin{equation}
\begin{gathered}
\tilde{E}^{(1)}_{1} = \int \mathcal{\tilde{H}}^{(1)}_{1} d^{3} x = \int \bigg\{ G_{0} [ f^{2}_{i} + f^{2} (\varepsilon_{ijk} \xi_{j} g_{k})^{2} ] + G_{1} \varepsilon_{ijk} \xi_{i} f_{j} g_{k} + \sum^{3}_{p=1} D_{p} G_{p+1} \bigg\} d^{3} x,
\end{gathered} 
\end{equation}
where $p=1,2,3$ and $G_{n} = G_{n}(f, g, \xi), (n = 0,...,4)$ are certain functions, which are to be determined later. 

The strong necessary conditions have the form:
\begin{eqnarray}
 \mathcal{\tilde{H}}^{(1)}_{1, f} &:& G_{0, f} [ f^{2}_{,i} + f^{2} (\varepsilon_{ijk} \xi_{j} g_{k})^{2} ] + 
 2 G_{0} f (\varepsilon_{ijk} \xi_{j} g_{k})^{2} + G_{1, f} \varepsilon_{ijk} \xi_{i} f_{j} g_{k} + \nonumber \\
 &+&
 \sum^{3}_{p=1} D_{p} G_{p+1, f} = 0, \nonumber \\ 
 \mathcal{\tilde{H}}^{(1)}_{1, g} &:& G_{0, g} [ f^{2}_{i} + f^{2} (\varepsilon_{ijk} \xi_{j} g_{k})^{2} ] + 
 G_{1, g} \varepsilon_{ijk} \xi_{i} f_{j} g_{k} + \sum^{3}_{p=1} D_{p} G_{p+1, g} = 0, \nonumber\\
 \mathcal{\tilde{H}}^{(1)}_{1, \xi} &:& G_{0, \xi} [ f^{2}_{i} + f^{2} (\varepsilon_{ijk} \xi_{j} g_{k})^{2} ] + 
  + G_{1, \xi} \varepsilon_{ijk} \xi_{i} f_{j} g_{k} + \sum^{3}_{p=1} D_{p} G_{p+1, \xi}= 0, \label{silne1}\nonumber \\
	\mathcal{\tilde{H}}^{(1)}_{1, f_{,r}} &:& 2 G_{0} f_{r} + G_{1} \varepsilon_{irk} \xi_{i} g_{k} + G_{r + 1, f} = 0, \nonumber \\
	\mathcal{\tilde{H}}^{(1)}_{1, g_{r}} &:& 2 G_{0} f^{2} \varepsilon_{mlr} \xi_{l} (\varepsilon_{ijk} \xi_{j} g_{k}) + 
	G_{1} \varepsilon_{ijr} \xi_{i} f_{j} + G_{r + 1, g} = 0,\nonumber \\
	\mathcal{\tilde{H}}^{(1)}_{1, \xi_{r}} &:& 2 G_{0} f^{2} \varepsilon_{mrl} g_{l} (\varepsilon_{ijk} \xi_{j} g_{k}) + 
	G_{1} \varepsilon_{rjk} f_{j} g_{k} + G_{r + 1, \xi} = 0, 
 \end{eqnarray}

 As usually in the case of strong necessary conditions, in order to derive Bogomolny decomposition (Bogomolny equations), 
 we need to make the equations (\ref{silne1}) to be self-consistent. This requires 
  \begin{eqnarray}
  G_{1} = 2 G_{0} f, \\
	G_{p+1} = const.,  p = 1, 2, 3, \\
	f_{i} - f \varepsilon_{ijk} \xi_{j} g_{k} = 0 .
	\end{eqnarray}	
	Then, three first equations are satisfied, and the Bogomolny decomposition has the form:
	\begin{equation}
	f_{i} - f \varepsilon_{ijk} \xi_{j} g_{k} = 0 . \label{BPS_pierwszy}
	\end{equation}
This Bogomolny equation can be used to find a topological bound on the energy of the first BPS submodel. Namely,
 \begin{eqnarray}
\hspace*{-1.0cm} E^{(1)}_{1} &=& 4 \int \frac{\sin^{2}{\xi}}{(1 + f^2)^{2}} \left[ f_{i} \pm f \varepsilon_{ijk} \xi_{j} g_{k} \right]^2 d^{3} x \mp  8 \int \frac{\sin^{2}{\xi}}{(1 + f^2)^{2}} \epsilon_{ijk}f f_i \xi_j g_k d^3 x \\
&\geq& \left| 8 \int \frac{\sin^{2}{\xi}}{(1 + f^2)^{2}} \epsilon_{ijk} f f_i \xi_j g_k d^3 x \right| = 4\pi^2 |B|
\end{eqnarray}
which is saturated if and only if the Bogomolny equation (\ref{BPS_pierwszy}) is obeyed. 

  \subsection{Second subsubmodel}

  Now we derive strong necessary conditions for (\ref{second_subsub}) with a generalization:

  \begin{equation}
  \begin{gathered}
   \tilde{E}^{(1)}_{2} = \int \mathcal{\tilde{H}}^{(1)}_{2} d^{3} x = 
	 \int \bigg \{ G_{0} [ f^{2} g^{2}_{i} + (\varepsilon_{ijk} \xi_{j} f_{k})^{2} ] 
   + G_{1} \varepsilon_{ijk} \xi_{i} f_{j} g_{k} + \sum^{3}_{p=1} D_{p} G_{p+1} \bigg \} d^{3} x 
  \end{gathered} 
  \end{equation}
  
	The strong necessary conditions have the form:
\begin{eqnarray}
 \mathcal{\tilde{H}}^{(1)}_{2, f} &:& G_{0, f} [ f^{2} g^{2}_{i} + (\varepsilon_{ijk} \xi_{j} f_{k})^{2} ] + 
 2 G_{0} f g^{2}_{k} + G_{1, f} \varepsilon_{ijk} \xi_{i} f_{j} g_{k}+ \nonumber \\ 
 &+& 
 \sum^{3}_{p=1} D_{p} G_{p+1, f} = 0, \nonumber \\
 \mathcal{\tilde{H}}^{(1)}_{2, g} &:& G_{0, g} [ f^{2} g^{2}_{i} + (\varepsilon_{ijk} \xi_{j} f_{k})^{2} ] + 
 G_{1, g} \varepsilon_{ijk} \xi_{i} f_{j} g_{k} + \sum^{3}_{p=1} D_{p} G_{p+1, g} = 0, \nonumber\\
 \mathcal{\tilde{H}}^{(1)}_{2, \xi} &:& G_{0, \xi} [ f^{2} g^{2}_{i} + (\varepsilon_{ijk} \xi_{j} f_{k})^{2} ] + 
  + G_{1, \xi} \varepsilon_{ijk} \xi_{i} f_{j} g_{k} + \sum^{3}_{p=1} D_{p} G_{p+1, \xi} = 0, \label{silne2} \nonumber \\
	\mathcal{\tilde{H}}^{(1)}_{2, f_{r}} &:& 2 G_{0} \varepsilon_{mlr} \xi_{l} (\varepsilon_{ijk} \xi_{j} f_{k}) + 
	G_{1} \varepsilon_{irk} \xi_{i} g_{k} + G_{r + 1, f} = 0,\nonumber \\
	\mathcal{\tilde{H}}^{(1)}_{2, g_{r}} &:& 2 G_{0} f^{2} g_{r} + G_{1} \varepsilon_{ijr} \xi_{i} f_{j} + G_{r + 1, g} = 0, \nonumber\\
	\mathcal{\tilde{H}}^{(1)}_{2, \xi_{r}} &:& 2 G_{0} \varepsilon_{mrl} f_{l} (\varepsilon_{ijk} \xi_{j} f_{k}) + 
	G_{1} \varepsilon_{rjk} f_{j} g_{k} + G_{r + 1, \xi} = 0, 
 \end{eqnarray}

 In order to make the equations (\ref{silne2}) self-consistent we have to put
  \begin{eqnarray}
  G_{1} = 2 G_{0} f, \\
	G_{p+1} = const.,  p = 1, 2, 3, \\
	f g_{i} + \varepsilon_{ijk} \xi_{j} f_{k} = 0 .
	\end{eqnarray}
  In this case, the Bogomolny decomposition has the form
	\begin{equation}
	f g_{i} + \varepsilon_{ijk} \xi_{j} f_{k} = 0 . \label{BPS_drugi}
	\end{equation} 
Corresponding topological bound on the energy reads 
 \begin{eqnarray}
\hspace*{-1.0cm} E^{(1)}_{2} &=& 4 \int \frac{\sin^{2}{\xi}}{(1 + f^2)^{2}} \left[ f g_{i} \pm \varepsilon_{ijk} \xi_{j} f_{k} \right]^2 d^{3} x \mp  8 \int \frac{\sin^{2}{\xi}}{(1 + f^2)^{2}} \epsilon_{ijk}f f_i \xi_j g_k d^3 x \\
&\geq& \left| 8 \int \frac{\sin^{2}{\xi}}{(1 + f^2)^{2}} \epsilon_{ijk} f f_i \xi_j g_k d^3 x \right| = 4\pi^2 |B|
\end{eqnarray}
which now is saturated if and only if the Bogomolny equation (\ref{BPS_drugi}) is obeyed. 

As we see the minimal Skyrme model $\mathcal{L}_{24}$ can be written as a sum of {\it three} BPS submodels. Each of them have the same topological bound $E \geq 4\pi |B|$, which however, is saturated for {\it different} field configurations as the corresponding Bogomolny equations are different. This provides a complete decomposition of the minimal Skyrme model as a sum of three coupled BPS submodels. Note also that none of them can be reached as a limit in the parameter space of the full model. 
	
	\section{Static solutions of the new BPS submodels}
	
 In order to understand solutions with a nontrivial topology in the upper defined BPS submodels we use the spherical coordinates and assume the following ansatz 
 \be
 \xi=\xi (r), \;\;\; f=f (\theta), \;\;\; g=n \phi
 \ee
 In addition one has to impose the usual boundary conditions which guarantee a nonzero baryon charge (the whole $\mathbb{S}^3$ target space must be covered at least once)
 \be
 \xi(r=0)=\pi, \;\; \xi(r=R)=0
 \ee 
 and
 \be
 f(\theta=0)= 0, \;\; f(\theta=\pi)=\infty
 \ee
 Here, $R$ is the geometric size of the soliton i.e., value of the radial coordinate where the profile function reaches the vacuum. 
 Then, the first Bogomolny equation (\ref{BPS_pierwszy}) leads to solutions
 \be
 \xi= \left\{
 \begin{array}{ll}
 \pi - C r & r \leq 1/C \\
 0 & r \geq 1/C
 \end{array} 
 \right.
 \ee
 and 
 \be
 f=A\left( \tan \frac{\theta}{2}\right)^{Cn}
 \ee
 where $C$ is a positive constant. Furthermore $A \in \mathbb{R}$. One can verify that such a solution has topological charge $B=n$. Note also that the size of solitons can be treated as a free parameter as
 \be
 R=1/C
 \ee
 
 The same ansatz aplied to the second BPS submodel gives
  \be
 \xi= \left\{
 \begin{array}{ll}
 \pi - D r & r \leq 1/D \\
 0 & r \geq 1/D
 \end{array} 
 \right.
 \ee
 and 
 \be
 f=A\left( \tan \frac{\theta}{2}\right)^{\frac{n}{D}}
 \ee
 Again, we find a one parameter family of compact solutions (compact Skyrmions) with topological charge $B=n$ and radius
 \be
 R=1/D
 \ee

 Several comments are in order. First of all, the obtained solutions of our two new BPS submodels $E^{(1)}_1$ and $E^{(1)}_2$ are, in generality, non-holomorphic configurations. Indeed, the angular part combines to holomorphic map only if $C=1$ or $D=1$, which is one of infinitely many possible solutions. Then,
 \be
 \omega = fe^{ig} = A\left( \tan \frac{\theta}{2}\right)^{n} e^{in \phi} = A z^n
 \ee
 It follows from this observation that, for $C=D=1$, these BPS submodels do have common solution which are exactly the compacton solution (with an arbitrary holomorphic map) of the $E^{(1)}$ BPS submodel \cite{submodels}. In other words, the holomorphic map solutions of the first BPS submodel $E^{(1)}$ emerge as a mutual effect of a competition of $E^{(1)}_1$ and $E^{(1)}_2$.
In addition, on the countrary to solutions of our submodels $E^{(1)}_1$ and $E^{(1)}_2$, solutions of $E^{(1)}$ have also a definite size. 

Next, the solutions of $E^{(1)}_1$ and $E^{(1)}_2$ are of the same (lower) type of continuity as compactons of $E^{(1)}$ submodel. Again, the first derivative of the profile is not continuous at the boundary while physical observables as energy density as well as topological charge density are continuos. 
\section{Summary}
In the present paper a complete decomposition of the minimal Skyrme model is performed. We have found that the model can be written as a sum of three BPS submodels with identical topological bounds.  These bounds are saturated if pertinent Bogomolny equations are obeyed, which are deferent for each submodel. Following that there are no common solutions as it should be since the minimal Skyrme model does not saturate the Faddeev bound. 

We also show how the rational maps (which are the main ingredient of the rational maps ansatz of the Skyrme model \cite{rational}) emerge due to a mutual interplay between new derived BPS submodels $E^{(1)}_1$ and $E^{(1)}_2$. 

On the other hand, the fact that each of the three BPS submodels $E^{(1)}_1$, $E^{(1)}_2$ and $E^{(2)}$ support also non-holomorphic BPS solutions may perhaps shed some lights on the role of non-holomorphic contribution of Skyrmions. In fact, it was observed that the rational map approximated solutions can be improved if a small non-holomorphic term is included \cite{krush}.

Finally, this hidden BPS structure of the full Skyrme model may be helpful in the construction of super-symmetric extensions of the Skyrme model \cite{susy}.

\section{Acknoweledgements}
  
  The author thanks to A. Wereszczynski for interesting discussions.

  \section{Computational resources}
  
     This research was supported in part by PL-Grid Infrastructure.

\end{document}